\documentclass[twocolumn,superscriptaddress,showpacs,nofootinbib,amsmath,amssymb,aps]{revtex4-1}

\usepackage{float}
\usepackage{graphicx}
\usepackage{dcolumn}
\usepackage{bm}
\usepackage{hyperref}
\usepackage{color}
\usepackage[caption=false]{subfig}
\usepackage{url}
\pdfoutput=1

\usepackage{graphicx}
\usepackage{dcolumn}
\usepackage{bm}
\usepackage[caption=false]{subfig}

\hypersetup{
	colorlinks = true,
	urlcolor   = blue,
	linkcolor  = blue,
	citecolor  = blue
}

\renewcommand*{\phi}{\varphi}
\renewcommand*{\epsilon}{\varepsilon}

\DeclareMathOperator{\Tr}{Tr}

\newcommand*{\bra}[1]{\langle #1 |}
\newcommand*{\ket}[1]{| #1 \rangle}

\newcommand*{\argmax}{\mathop{\mathrm{argmax}}}

\newcommand*{\prob}{\mathbb P}

\begin{document}

\title{Experimental neural network enhanced quantum tomography}\thanks{All data and source code are available online at \cite{code}.}

\author{Adriano Macarone Palmieri}
\affiliation{Deep Quantum Labs, Skolkovo Institute of Science and Technology, Moscow 121205, Russia}

\author{Egor Kovlakov}
\affiliation{Quantum Technologies Centre, Faculty of Physics, M.~V.~Lomonosov Moscow State University, Moscow 119991, Russia}

\author{Federico Bianchi}
\affiliation{Department of Computer Science, University of Milan-Bicocca, Milan 20133, Italy}

\author{Dmitry Yudin}
\affiliation{Deep Quantum Labs, Skolkovo Institute of Science and Technology, Moscow 121205, Russia}

\author{Stanislav Straupe}
\affiliation{Quantum Technologies Centre, Faculty of Physics, M.~V.~Lomonosov Moscow State University, Moscow 119991, Russia}

\author{Jacob D.~Biamonte}
\affiliation{Deep Quantum Labs, Skolkovo Institute of Science and Technology, Moscow 121205, Russia}

\author{Sergei Kulik}
\affiliation{Quantum Technologies Centre, Faculty of Physics, M.~V.~Lomonosov Moscow State University, Moscow 119991, Russia}


\begin{abstract}
Quantum tomography is currently ubiquitous for testing any  implementation of a quantum information processing device. Various sophisticated procedures for state and process reconstruction from measured data are well developed and benefit from precise knowledge of the model describing state preparation and the measurement apparatus. However, physical models suffer from intrinsic limitations as actual measurement operators and trial states cannot be known precisely. This scenario inevitably leads to state-preparation-and-measurement (SPAM) errors degrading reconstruction performance. Here we develop and experimentally implement a machine learning based protocol reducing SPAM errors. We trained a supervised neural network to filter the experimental data and hence uncovered salient patterns that characterize the measurement probabilities for the original state and the ideal experimental apparatus free from SPAM errors. We compared the neural network state reconstruction protocol with a protocol treating SPAM errors by process tomography, as well as to a SPAM-agnostic protocol with idealized measurements. The average reconstruction fidelity is shown to be enhanced by 10\% and 27\%, respectively.  The presented methods apply to the vast range of quantum experiments which rely on tomography. 
\end{abstract}

\maketitle

{\it Introduction.}~Rapid experimental progress realizing quantum enhanced technologies places an increased demand on methods for validation and testing. As such, various approaches to augment state- and process-tomography have recently been proposed. A persistent problem faced by these contemporary approaches are systematic errors in state preparation and measurements (SPAM).  Such notoriously challenging errors are inevitable in any experimental realization \cite{Rosset_PRA2012,Knill_PRA2008,Merkel_PRA2013,Blume_NatCom2017,Lundeen_NatPhys2009,Brida_PRL2012,Houlsby_PRA12,Kravtsov_PRA13,Torlai_NatPhys2018,CarloDBM,Rocchetto_npjQI2018}.  Here we develop a data-driven, deep-learning based approach to augment state- and detector-tomography that successfully minimized SPAM error on quantum optics experimental data. 

Several prior approaches have been developed to circumvent the SPAM problem. One line of thought leads to the so-called \textit{randomized benchmarking} protocols \cite{Knill_PRA2008,Emerson_PRL2011,Wallman_NJP2014}, which were designed for quality estimation of quantum gates in the quantum circuit model. The idea is to average the error over a large set of randomly chosen gates, thus effectively minimizing the average influence of SPAM. Randomized benchmarking in its initial form however, only allowed to estimate an average fidelity for the set of gates, so more elaborate and informative procedures were developed \cite{Merkel_PRA2013,Roth_PRL2018}. Another example is \textit{gate set tomography} \cite{Blume_ArXiv2013,Blume_NatCom2017,Dehollain_NJP2016}.  Therein the experimental apparatus is treated as a black box with external controls allowing for (i) state preparation, (ii) application of gates and (iii) measurement.  These unknown components (i)-(iii) are inferred from measurement statistics. Both approaches require long sequences of gates and are not suited for a simple prepare-and-measure scenario in quantum communication applications. Indeed, in such a scenario the experimenter faces careful calibration of the measurement setup, or in other words \textit{quantum detector tomography} \cite{Lundeen_NatPhys2009,Bobrov_OpEx2015,Brida_PRL2012}, which works reliably if known probe states can be prepared \cite{Mogilevtsev_NJP2012,Bra_czyk_NJP2012,Straupe_PRA2013,Jackson_PRA2015}.

As (imperfect) quantum tomography is a data-driven technique, recent proposals suggest a natural benefit offered by machine learning methods.  Bayesian models were used to optimise the data collection process by adaptive measurements in state reconstruction \cite{Houlsby_PRA12,Kravtsov_PRA13,Granade_NJP2017}, process tomography \cite{Kulik_PRA17}, Hamiltonian learning \cite{Granade_NJP12} and other problems in experimental characterisation of quantum devices \cite{Lennon_ArXiv2018}. Neural networks were proposed to facilitate quantum tomography in high-dimensions. In such approaches neural networks of different architectures, such as restricted Boltzmann machines \cite{Torlai_NatPhys2018,Carrasquilla_NMI2019, CarloDBM}, variational autoencoders \cite{Rocchetto_npjQI2018} and other architectures \cite{Xin_ArXiv2018} are used for efficient state reconstruction; interestingly, a model for tackling a more realistic scenario of mixed quantum states has been proposed \cite{latentSpacePur}. 

Our framework differs significantly and is based on supervised learning, specifically tailored to address SPAM errors.  Our method hence compensates for measurement errors of the specific experimental apparatus employed, as we demonstrate on real experimental data from high-dimensional quantum states of single photons encoded in spatial modes. The success of our approach bootstraps the well-known \textit{noise filtering} class of techniques in machine learning.  

{\it Quantum tomography.} Performing quantum state estimation implies the reconstruction of the density matrix $\rho$ of an unknown quantum state given the outcomes of known measurements \cite{Banaszek_PRA99,James01,Paris_Book2004}. In general, a measurement is characterized by a set of positive operator valued measures (POVM's) $\{\mathbb M_a\}$ with index $\alpha \in \mathcal A$ the different configurations of the experimental apparatus (set $\mathcal A $). Given the configuration $\alpha$, the probability of observing an outcome $\gamma$ is given:
\begin{equation}\label{eq:BornRule}
\prob(\gamma | \alpha, \rho) = \Tr(M_{\alpha\gamma} \rho),
\end{equation}
where $M_{\alpha\gamma} \in \mathbb M_\alpha$~ are POVM elements, i.e. positive operators satisfying the completeness relation $\sum_\gamma M_{\alpha\gamma} = \mathbb{I}$. A statistical estimator maps the set of all observed outcomes $\mathcal D_N = \{\gamma_n\}_{n=1}^N$ onto an estimate of the unknown quantum state $\hat \rho$. A more general concept of quantum process tomography stands for a protocol dealing with estimation of an unknown quantum operation acting on quantum states \cite{Chuang1997,Poyatos_PRL97}. Process tomography uses measurements on a set of known test states $\{\rho_\alpha\}$ to recover the description of an unknown operation.\footnote{See Supplemental Material (Section~\ref{app:Gouy}) for the thorough discussion of quantum process tomography and its application for calibration of the measurement setup (Section~\ref{app:experiment}).}

The reconstruction procedure requires knowledge of the measurement operators $\{M_{\alpha\gamma}\}$, as well as the test states $\{\rho_\alpha\}$ in the case of process tomography. However, both tend to deviate from the experimenter's expectations due to stochastic noise and systematic errors. While stochastic noise may to some extent be circumvented by increasing the sample size, systematic errors are notoriously hard to correct. The only known way to make tomography reliable is to explicitly incorporate these errors in (\ref{eq:BornRule}). Thus, trial states and measurements should be considered as acted upon by some SPAM processes: $\tilde\rho_\alpha=\mathcal{R}(\rho_\alpha)$ and $\tilde M_{\alpha\gamma}=\mathcal{M}(M_{\alpha\gamma})$, and the models for these processes should be learned independently from a calibration procedure. Such calibration is essentially tomography on its right. For example, the reconstruction of measurement operators is known as detector tomography \cite{Fiurasek_PRA01,Lundeen_NatPhys2009,Bobrov_OpEx2015,Brida_PRL2012,Maccone_PRL04} and requires ideal preparation of calibration states. The most straightforward approach is calibration of the measurement setup with some close-to-ideal and easy to prepare test states, or calibration of the preparation setup with known and close-to-ideal measurements. In this case, one may then infer the processes $\mathcal{R}$ and/or $\mathcal{M}$ explicitly -- for example -- in the form of the corresponding operator elements, and incorporate this knowledge in the reconstruction procedure. Ideally, this procedure should produce an estimator free from bias caused by systematic SPAM errors.\footnote{See Supplemental Material (Section~\ref{app:Gouy}) for the detailed description of this procedure applied to our experiment (Section~\ref{app:generation}).}

{\it Denoising by deep learning.} The problem of fighting SPAM is essentially a denoising problem. Given the estimates of raw probabilities inferred from the experimental dataset $\tilde\prob(\gamma | \alpha, \tilde\rho) = \Tr\small(\tilde M_{\alpha\gamma}\tilde\rho\small),$ one wants to establish a one-to-one correspondence with the ideal probabilities $\tilde\prob(\gamma | \alpha, \tilde\rho) \leftrightarrow \prob(\gamma | \alpha, \rho)$ for the measurement setup free from systematic SPAM errors. We use a deep neural network (DNN) in the form of an overcomplete autoencoder trained on a dataset $\mathcal D_N$ to approximate the map from $\tilde\prob$ to $\prob$. 

To train and test the DNN we prepare a dataset of $N$ Haar-random pure states $\mathcal D_N =\{\ket{\psi_i}\}_{i=1}^N$. For a $d$-dimensional Hilbert space, reconstruction of a Hermitian density matrix with unit trace requires at least $d^2$ different measurements. The network is trained on the dataset, consisting of $d^2 \times N$ frequencies experimentally obtained by performing the same $d^2$ measurements $\{\tilde M_\gamma\}_{\gamma=1}^{d^2}$ for all $N$ states (in our experiments $d=6$, i.e. we deal with a six-dimensional Hilbert space). These frequencies are fed to the input layer of the feed-forward network consisting of $d^2=36$ neurons.\footnote{See Supplemental Material (Section~\ref{app:nn}) for DNN architecture and the details of training process.}

\begin{figure}
    \centering
    \includegraphics[width = 0.5\textwidth]{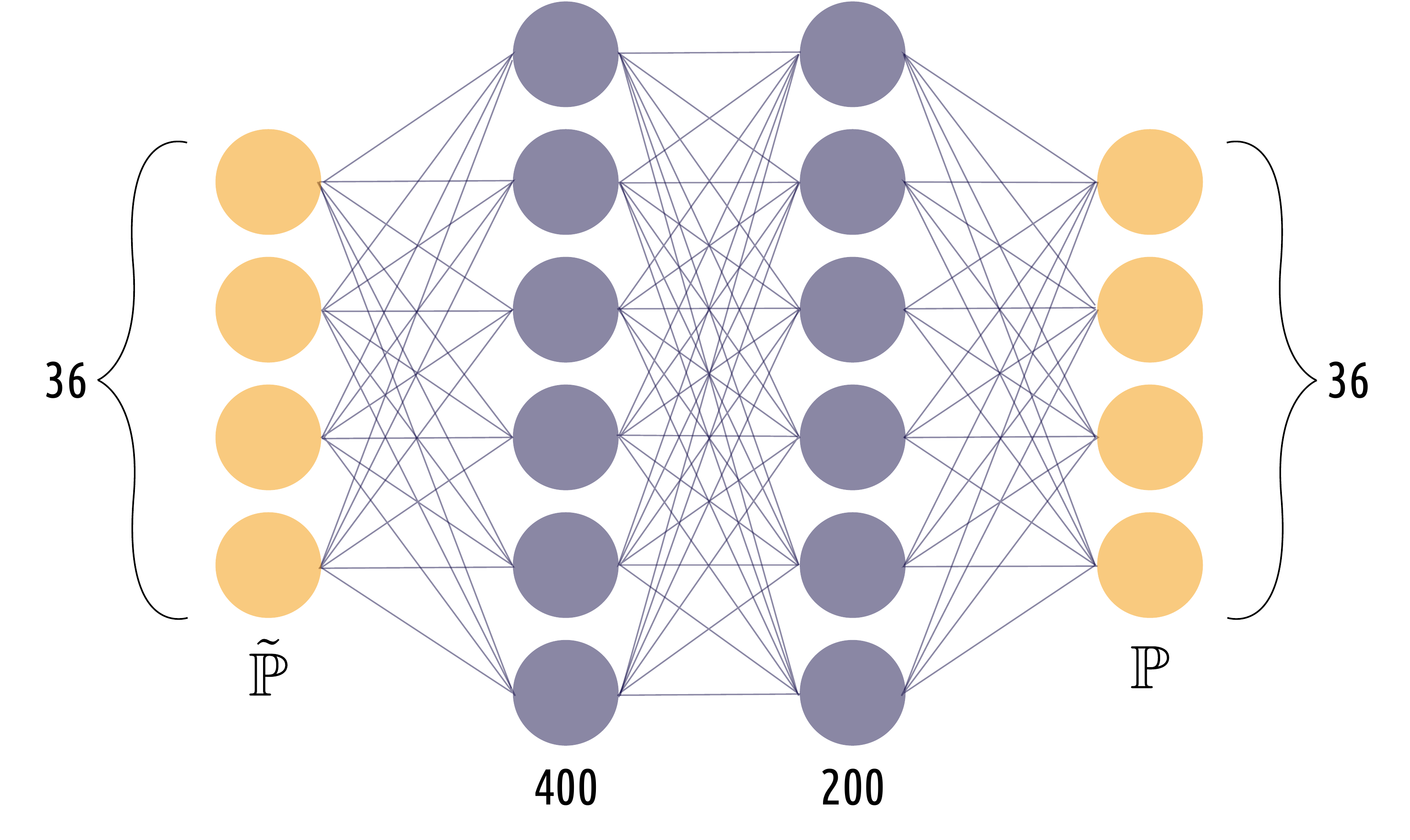}
    \caption{The DNN architecture for an overcomplete autoencoder, employed in our simulations for denoising. Input and output layers constitute of 36 neurons each and two hidden layers of 400 and 200 neurons respectively. The DNN modifies its internal parameters to find a function $\mathcal{F}:\tilde\prob(\gamma | \alpha, \tilde\rho)\rightarrow\prob(\gamma | \alpha, \rho)$ which translates between the experimentally estimated probabilities $\tilde\prob(\gamma | \alpha, \tilde\rho)$, subjected to SPAM errors, at the input and ideal $\prob(\gamma | \alpha, \rho)$ at the output. To achieve this goal the network is forced to reduce the Kullback-Leibler divergence amongst pairs of distributions. An early stopper is applied in order to avoid overfitting during the training phase.}
    \label{fig:dnn}
 \end{figure}

We use a DNN with two hidden layers as shown in Fig.~\ref{fig:dnn}. The first hidden layer is chosen to consist of four hundred neurons, whilst the second contains two hundred. To prevent overfitting we applied dropout between the two hidden layers with drop probability equal to 0.2, i.e.~at each iteration we randomly drop 20\% neurons of the first hidden layer in such a way that the network becomes more robust to variations. We use a {\it rectified linear unit} as an activation function after both hidden layers, while in the final output $d^2$-dimensional layer we use a {\it softmax} function to transform the predicted values to valid normalized probability distributions. Following the standard paradigm of statistical learning, we divided our dataset of overall $N=10500$ states (represented by their density matrix elements) into $7000$ states for training, $1500$ states for validation and $2000$ for testing. The validation set is an independent set and is used to stop the network training as soon as the error evaluated for this set stops decreasing (generally, this is referred to as early stopping: we examine validation loss every 100 epochs). Our loss function is computed over mini-batches of data of size 40.

{\it Kullback-Leibler divergence.} Training is performed by minimization of the loss function, defined as the sum of Kullback-Leibler divergences between the distributions of \emph{predicted probabilities} $\{p^i_\gamma\}_{\gamma=1}^{d^2}$ at the output layer of the network and the ideally \textit{expected probabilities} $\{\prob^i_\gamma\}_{\gamma=1}^{d^2}$, which are calculated for the test states as $\prob^i_\gamma= \Tr(M_{\gamma} \rho_i)$ assuming errorless projectors $M_{\gamma}$: 
\begin{equation}\label{eq:KL}
	L=\sum_{i=1}^{N} D_{KL}(\{\prob^i\}||\{p^i\}) = \sum_{i=1}^{N} \sum_{\gamma=1}^{d^2} \prob^i_\gamma \log\left( \frac{\prob^i_\gamma}{p^i_\gamma} \right).
\end{equation}
The minimization of KL divergence of Eq.~(\ref{eq:KL}) is achieved by virtue of gradient descent with respect to the parameters $\{\theta_k\}$ of the DNN for updating its internal weights. The KL divergence for a pair $(\prob^i,p^i)$ can be expressed in terms of cross-entropy $H(\prob^i,p^i)=\sum_{\gamma=1}^{d^2}\prob^i_\gamma\log p^i_\gamma$ which has to be minimized. For this purpose, we utilized the {\it RMSprop} \cite{rmsp} algorithm, in which the learning rate is adapted for each of the parameters $\{\theta_k\}$, dividing the learning rate for a weight by a running average of the magnitudes of recent gradients for that weight according to
 \begin{align}\label{rprop}
    v_t=\alpha v_{t-1}+(1-\alpha)(\nabla_i L^2(\{\theta\}))
 \end{align}
where $\alpha = 0.1$. While the parameters are updated as
\begin{equation}\label{rprop1}
    \theta_t=\theta_{t-1}- \tfrac{\eta}{\sqrt{v(t)}}\nabla_i L(\{\theta\})
\end{equation}
with $\eta$ standing for the learning rate.

{\it Experimental dataset.} We fix the set of tomographicaly complete measurements $\{\mathbb M_\alpha\}=\mathbb M$ to estimate all matrix elements of $\rho$ using (\ref{eq:BornRule}) and an appropriate estimator. We will assume that our POVM $\mathbb M$ consists of $d^2$ one-dimensional projectors $M_\gamma = \ket{\varphi_\gamma}\bra{\varphi_\gamma}$. These projectors are transformed by systematic SPAM errors into some positive operators $\tilde M_\gamma$. Experimental data consists of frequencies $f_\gamma=n_\gamma/n$, where $n_\gamma$ is the number of times an outcome $\gamma$ was observed in a series of $n$ measurements with identically prepared state $\rho$. For the time being, we assume, that all the SPAM errors can be attributed to the measurement part of the setup, and the state preparation may be performed reliably. This is indeed the case in our experimental implementation (see Supplemental Material). 

\begin{figure}
	\center{\includegraphics[width=\linewidth]{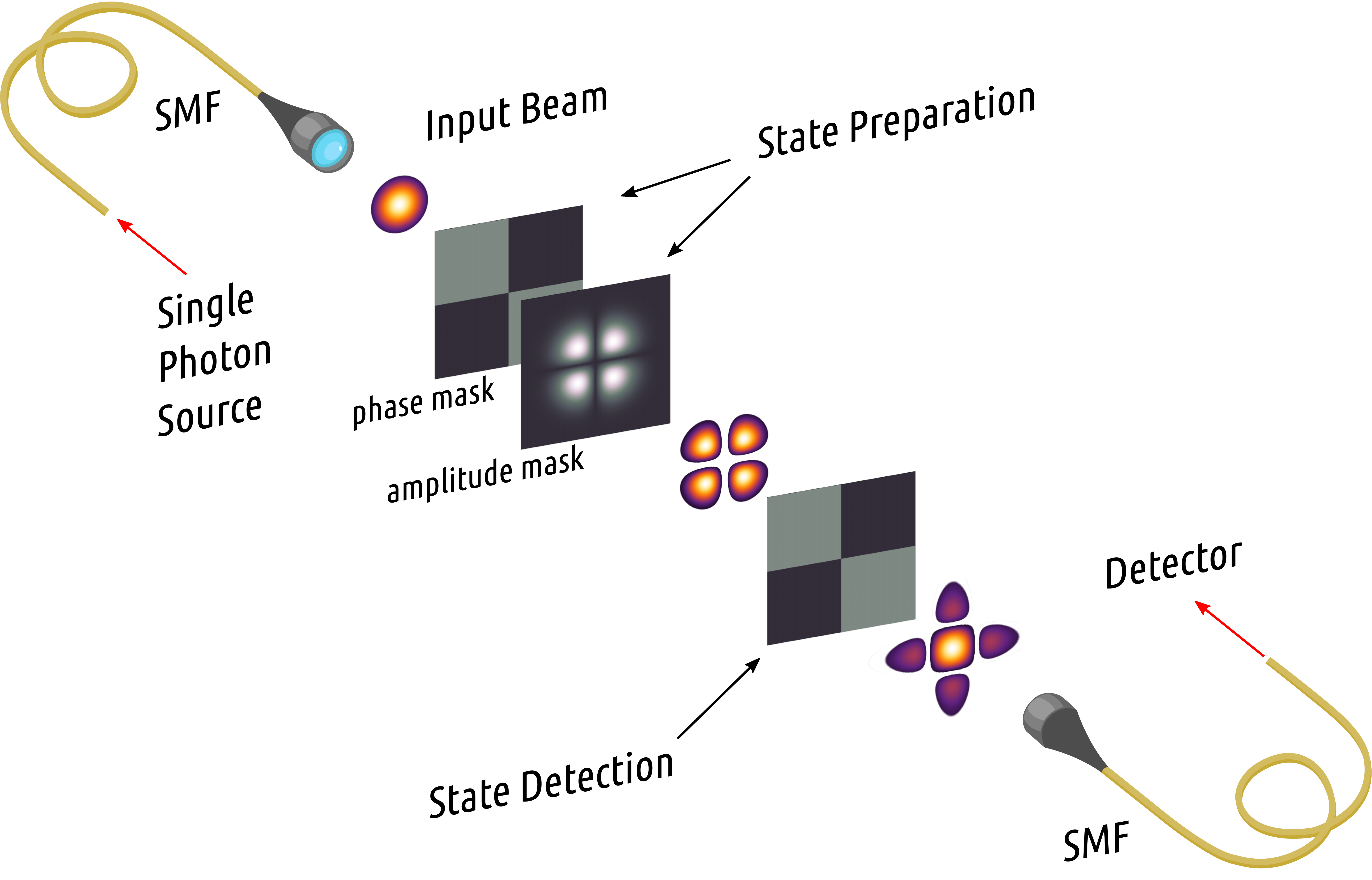}}
	\caption{Experimental setup for preparation and measurement of spatial qudit states. In the generation part, single photons from a heralded source are beam-shaped by a single mode fiber (SMF) and then transformed by a hologram displayed on a spatial light modulator. Analogously, the detection part consists of a hologram corresponding to the chosen detection mode, followed by a single mode fiber and a single photon counter. The hologram in the generation part produces high-quality HG modes with the use of amplitude modulation, while a phase-only hologram at the detection part sacrifices projection quality for efficiency.}\label{fig:setup}
\end{figure}

We reconstruct high-dimensional quantum states encoded in the spatial degrees of freedom of photons. The most prominent example of such encoding uses photonic states with orbital angular momentum (OAM) \cite{Molina-Terriza_NPhys2007} as relevant to numerous experiments in quantum optics and quantum information. However, OAM is only one of two quantum numbers, associated with orthogonal optical modes, and radial degree of freedom of Laguerre-Gaussian beams \cite{Salakhutdinov_PRL2012,Krenn_PNAS2014} as well as full set of Hermite-Gaussian (HG) modes \cite{Kovlakov_PRL2017} offer viable alternatives for increasing the accessible Hilbert space dimensionality. One of the troubles with using the full set of orthogonal modes for encoding is the poor quality of projective measurements. Existing methods to remedy the situation \cite{bouchard2018measuring} trade reconstruction quality for efficiency, significantly reducing the latter. Complex high-dimensional projectors are especially vulnerable to measurement errors and fidelities of state reconstruction are typically at most $\sim 0.9$ in high-dimensional tomographic experiments \cite{bent2015experimental}. That provides a challenging experimental scenario for our machine-learning-enhanced methods.

\begin{figure*}[hbt!]
	\subfloat[Fidelity]
	{	
		\includegraphics[width=0.3\linewidth]{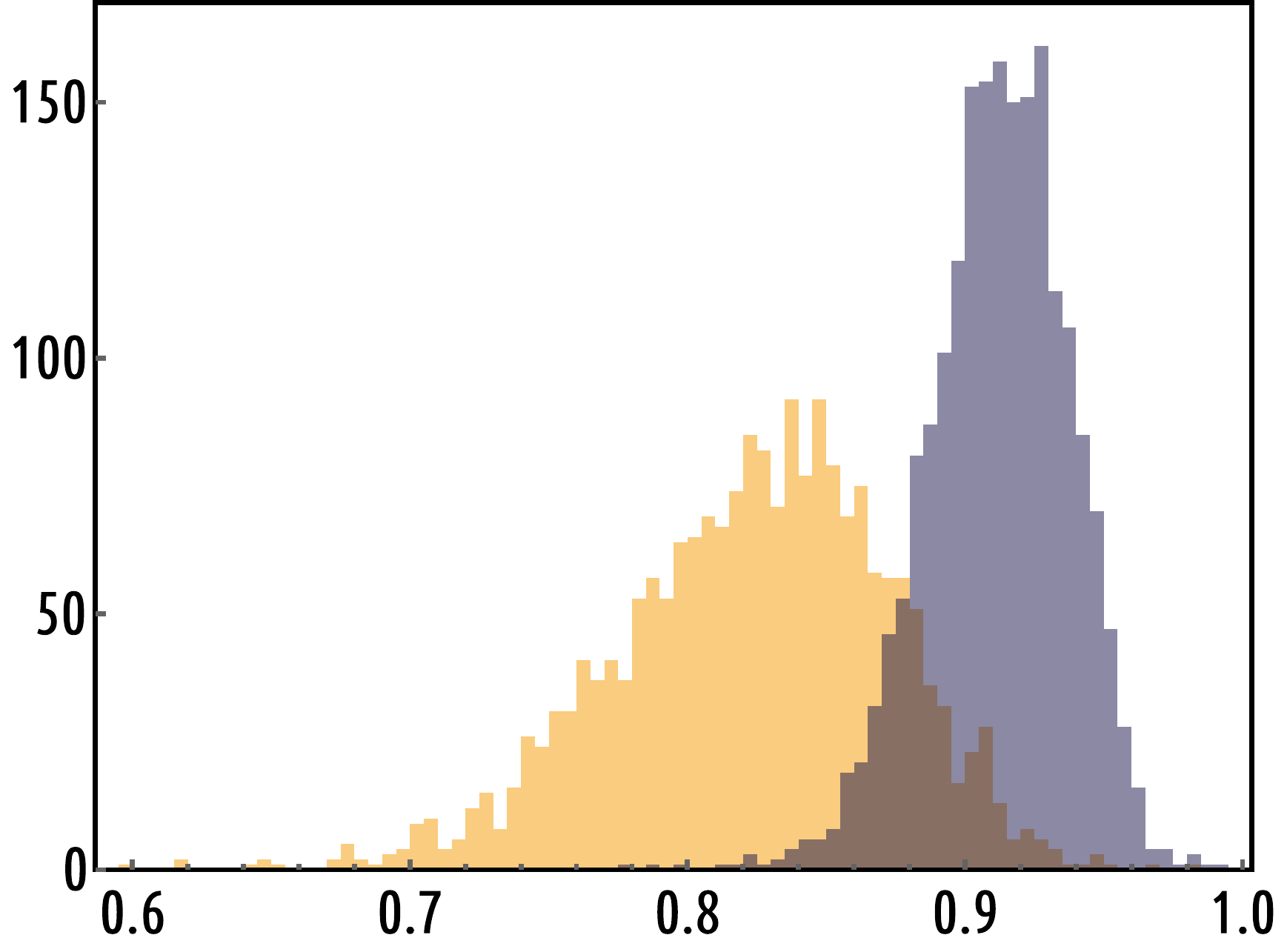}
		\label{fig:fidelity}
	}
	\subfloat[Purity]
	{
		\includegraphics[width=0.3\linewidth]{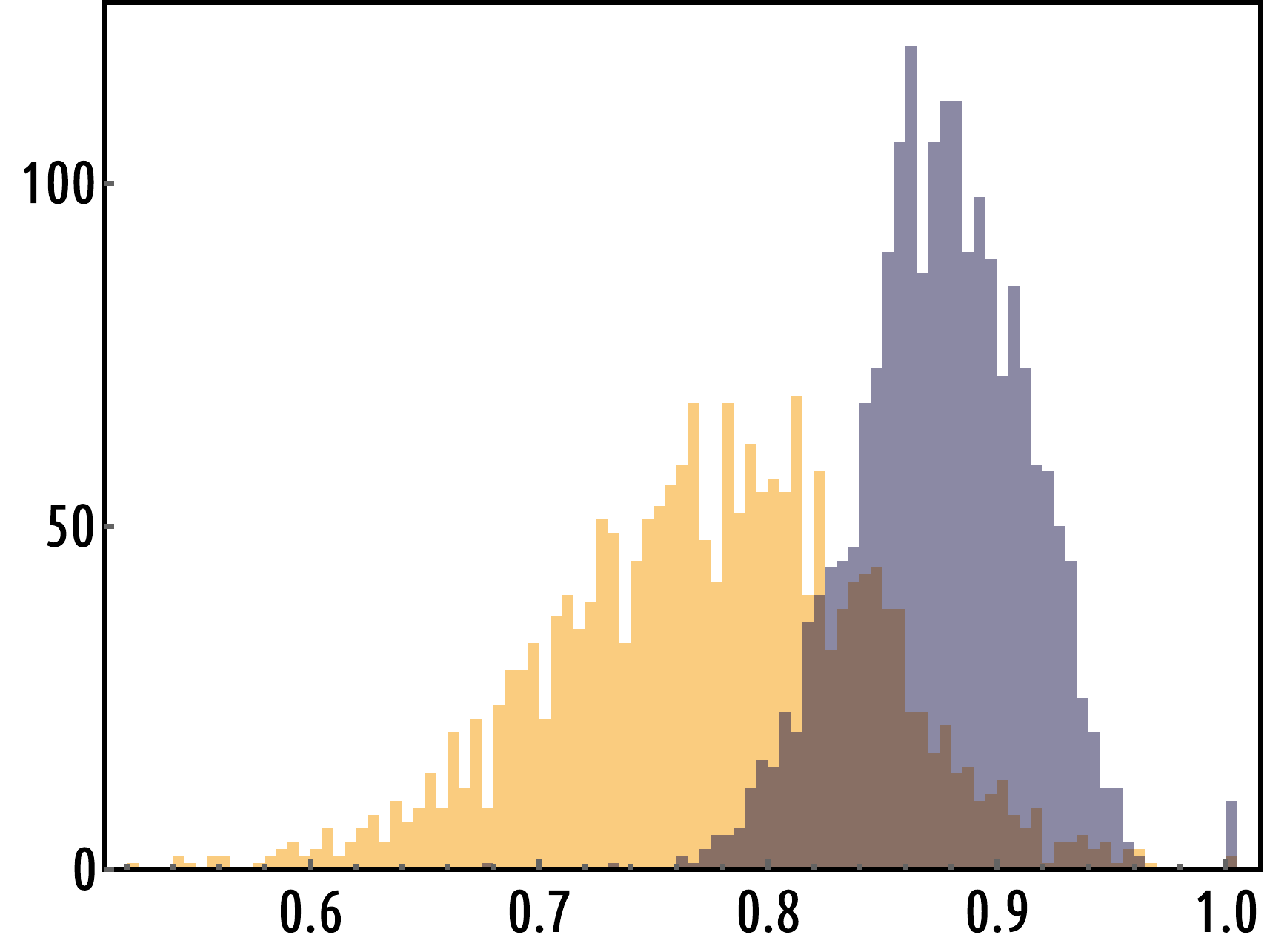}
		\label{fig:purity}
	}
	\subfloat[Fidelity (pure)]
	{
		\includegraphics[width=0.3\linewidth]{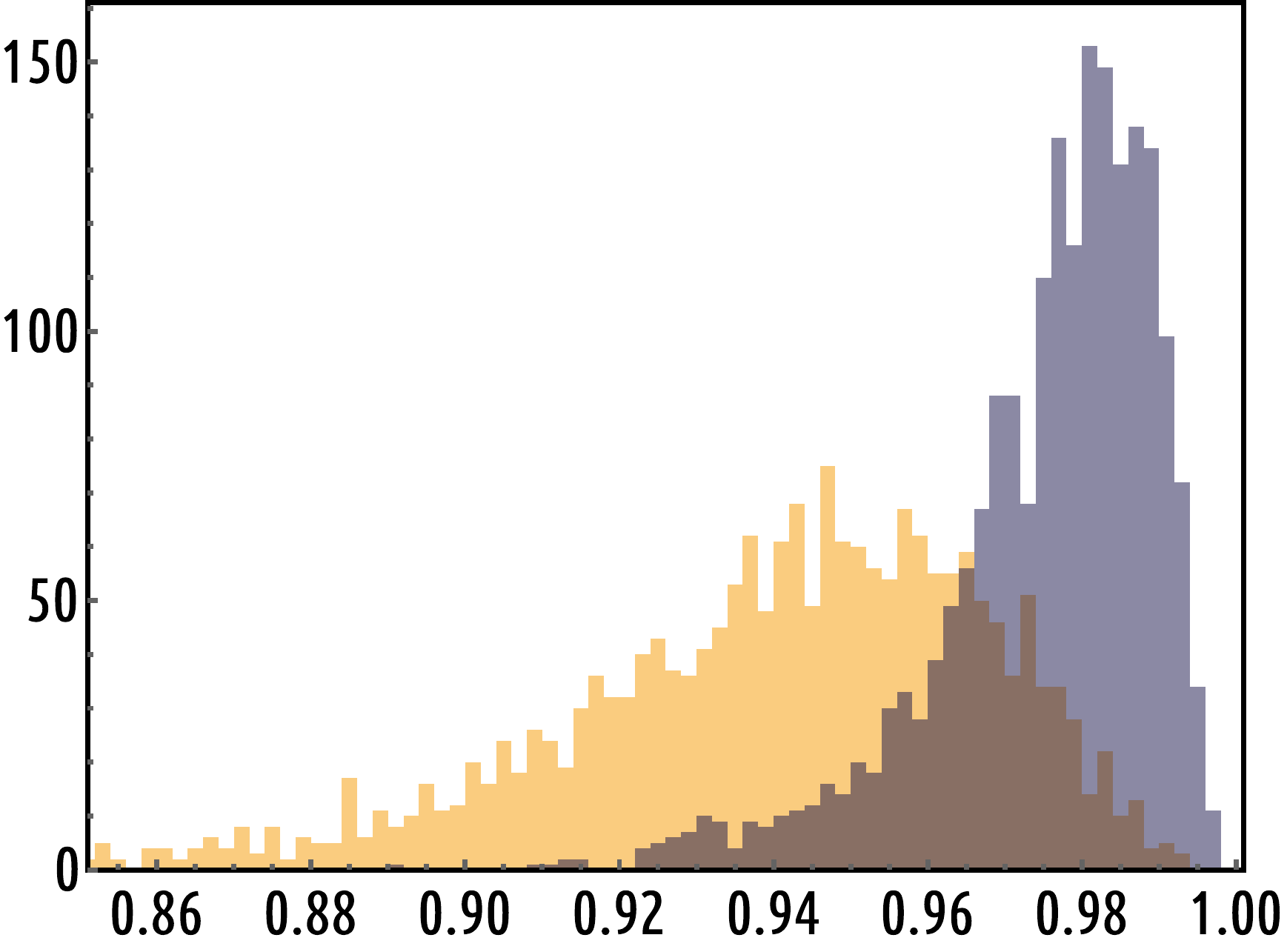}
		\label{fig:fidelityPure}
	}
	\caption{Results of experimental state reconstruction vs phase only holograms. (a) Fidelity of the experimentally reconstructed states with ideal  $F=\bra{\psi^i}\hat{\rho}^i_{(raw/nn)}\ket{\psi^i}$ for 2000 test states reconstructed from raw data (orange bars) and reconstructed after neural network processing of the data (blue bars). (b) A similar diagram for purity of the reconstructed states, $\pi=\Tr \hat\rho^2$. (c) Fidelity histogram for the case, when the state is reconstructed to be pure. The results of the filtering process are clearly witnessed by the modification of data histogram shapes. Besides the shifting towards higher values, that shows average gain over our experimental data, the reduction of FWHM indicates filtering task by the neural network.}
	\label{fig:ExpReconstruction}
\end{figure*}

Our experiment is schematically illustrated in Fig.~\ref{fig:setup}. We use phase holograms displayed on the spatial light modulator as spatial mode transformers. At the preparation stage an initially Gaussian beam is modulated both in phase and in amplitude to an arbitrary superposition of HG modes, which are chosen as the basis in the Hilbert space. At the detection phase the beam passes through a phase-only mode-transforming hologram and is focused to a single mode fiber, filtering out a single Gaussian mode. This sequence corresponds to a projective measurement in mode space, where the projector $\tilde{M}_\gamma$ is determined by the phase hologram.\footnote{See Supplemental Material for the details of the experimental setup (Section~\ref{app:experiment}) and state preparation and detection methods (Section~\ref{app:generation}).} In dimension $d=6$, we are able to prepare an arbitrary superposition expressed in the basis of HG modes as $\ket{\psi}=\sum_{i,j=0}^{2}c_{ij}\ket{\mathrm{HG}_{ij}}$. In the measurement phase we used a SIC (symmetric informationally complete) POVM, which is close to optimal for state reconstruction and may be relatively easily realized for spatial modes \cite{bent2015experimental}. 

{\it Experimental results.} We performed state reconstruction using maximum likelihood estimation \cite{Hradil97} for both raw experimental data and DNN-processed data.\footnote{See also Supplemental Material (Section~\ref{app:Data}) for extra information on spatial probability distribution of reconstructed states.} In the former case, the log-likelihood function to be maximized with respect to $\rho$ has been chosen as $\mathcal{L}(f^i_\gamma|\rho)\propto\sum_{\gamma=1}^{36}f^i_\gamma\log\left[\Tr (M_\gamma \rho)\right]$, with frequencies $f_\gamma=n_\gamma/n$ and $i$ numbering the test set states. Whereas in the latter case, these frequencies have been replaced with predicted probabilities $p_\gamma$. The results for $\hat\rho^i_{(raw)} = \argmax \mathcal{L}(f^i_\gamma|\rho)$ and $\hat\rho^i_{(nn)} = \argmax \mathcal{L}(p^i_\gamma|\rho)$ with the prepared states $\ket{\psi^i}$ are shown in Fig.~\ref{fig:ExpReconstruction}.~Interestingly, the average reconstruction fidelity increases from $F_{(raw)}=(0.82\pm0.05)$ to $F_{(nn)}=(0.91\pm0.03)$ and this increase is uniform over the entire test set. Similar behavior is observed for the purity --- since we did not force the state to be pure in the reconstruction, the average purity of the estimate is less then unity: {$\pi_{(raw)}=(0.78\pm0.07)$, whereas $\pi_{(nn)}=(0.88\pm0.04)$}. If the restriction to pure states is explicitly imposed in the reconstruction procedure, the fidelity increase is even more significant, as shown in Fig.~\ref{fig:fidelityPure}. In this case the initially relatively high fidelity of ${F}_{(raw)}=(0.94\pm0.03)$ increases to ${F}_{(nn)}=(0.98\pm0.02)$ --- a very high value, given the states dimensionality.

{\it Conclusion.} Our results  were obtained with analytical correction for some known SPAM errors already performed. In particular, we have explicitly taken into account the Gouy phase-shifts acquired by the modes of different order during propagation {(see Supplemental Material)}. This correction is however unnecessary for neural-network post-processing. The DNN has been trained without any need of data \textit{preprocessing} over the experimental dataset, as to say  without introducing any phase correction in our initial data, wherein considering the effect of a channel process $\mathcal{E}$. However, we have achieved average estimation fidelities of $F_{(nn)}= (0.81\pm0.19)$ as compared to $F_{(raw)}=(0.54\pm0.12)$ for this \textit{completely agnostic} scenario, showing a dramatic improvement by straightforward application of a learning approach. To conclude, our results unambiguously demonstrate that a use of neural-network-architecture on experimental data can provide a reliable tool for quantum state-and-detector tomography.

{\it Acknowledgements.} The authors acknowledge financial support under the Russian National Quantum Technologies Initiative and thank Timur Tlyachev and Dmitry Dylov for helpful suggestions on an early version of this study, and Yuliia Savenko (illustrator) for producing the neural network illustration. E.K. acknowledges support from the BASIS Foundation.

\newpage
\appendix 

\onecolumngrid

\section*{Supplemental Material}

\vskip 10pt

\subsection{Experimental setup}\label{app:experiment}
We use spatial degrees of freedom of photons to produce high-dimensional quantum states. The corresponding continuous Hilbert space is typically discretized using the basis of transverse modes, for this purpose we chose Hermite-Gaussian (HG) modes $\mathrm{HG}_{nm}(x,y)$, which are the solutions of the Helmholtz equation in Cartesian coordinates $(x,y)$ and form a complete orthonormal basis. The HG modes are separable in $x$- and $y$-coordinates, so that $\mathrm{HG}_{nm}(x,y) = \mathrm{HG}_{n}(x) \times \mathrm{HG}_{m}(y)$. Each mode is characterized by indices $n$ and $m$ which indicate the orders of corresponding Hermite polynomials $H_{n}(x)$ and $H_{m}(y)$:
\begin{equation}\label{eq:HGmode}
\mathrm{HG}_{n}(z) \propto H_{n}(\sqrt{2}z/w) \exp (-z^{2} / w^{2}),
\end{equation}
where $w$ is the mode waist. We limited the dimensionality of the Hilbert space to 6 by using only the beams with $n+m \leq 2$. The basis of HG modes is fully equivalent to a commonly used basis of Laguerre-Gaussian (LG) modes which are also the solutions of the Helmholtz equation but in cylindrical coordinates. Most commonly, only the azimuthal part of LG basis, associated with orbital angular momentum (OAM) of photons, is considered in the experiments, primarily due to simplicity of detection \cite{bent2015experimental}. Here we use the full two-dimensional mode spectrum of HG modes, which is equivalent to including the radial degree of freedom in addition to OAM. This is rarely done in quantum experiments, and one of the reasons is poor quality of projective measurements. Thus, this choice of physical system nicely fits the purpose of our demonstration.

\begin{figure} [h]
	\center{\includegraphics[width=0.6\linewidth]{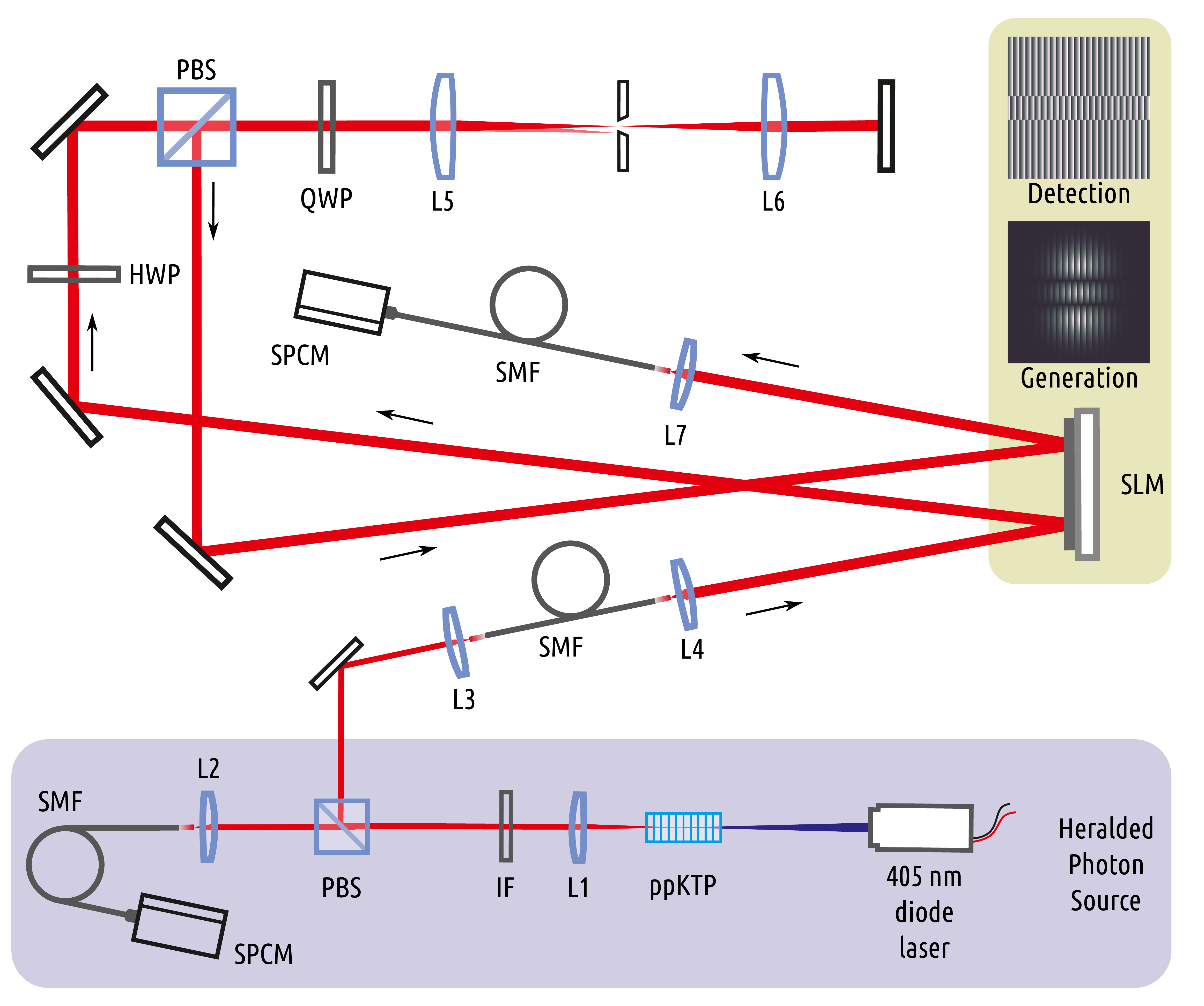}}
	\caption{Experimental setup for preparation and measurement of spatial qudit states. In the generation part single photons from a heralded source or a diode laser at the same wavelength of 810~nm are beam-shaped by a single mode fiber (SMF) and then transformed by a hologram displayed on a spatial light modulator. The detection part analogously consists of a hologram, corresponding to the chosen detection mode followed by a single mode fiber and a single photon counter. The hologram in the generation part uses amplitude modulation to produce high-quality HG modes, while a phase-only hologram at the detection part sacrifices projection quality for efficiency.}\label{fig:scheme}
\end{figure}
 
The experimental setup is presented in Fig.~\ref{fig:scheme}. Two light sources were used: an attenuated 808~nm diode laser and a heralded single photon source. Heralded single photons were obtained from spontaneous parametric down conversion in a 15 mm periodically-poled KTP crystal pumped by a 405~nm volume-Bragg-grating-stabilized diode laser. Beams from both sources were filtered by a single-mode fiber (SMF) and then collimated by an aspheric lens L2 (11 mm). We used one half (right) of the SLM (Holoeye Pluto) to generate the desired mode in the first diffraction order of the displayed hologram. Since SLM's working polarization is vertical, the half-wave plate (HWP) was inserted into the optical path to let the beam pass through the polarizing beamsplitter (PBS). The combination of lenses L3 and L4 with equal focal lengths (100 mm) separated by a 200 mm distance was used to cut off the zero diffraction order with a pinhole in the focal plane. After the double pass through this telescope and a quarter-wave plate (QWP) the beam was reflected by the PBS and directed back to the SLM. Using the hologram displayed on the left half of the SLM and a single mode fiber followed by a single photon counting module (SPCM) we realized a well-known technique of projective measurements in the spatial mode space \cite{mair2001entanglement}. To focus the first diffraction order of the reflected beam on the tip of the fiber we used an aspheric lens L5 with the same focal length (11 mm) as L1.

All data used for the neural network training and evaluation were taken for an attenuated laser source due to much higher data acquisition rate. When the NN trained on the attenuated laser was applied to a dataset taken with the heralded single photon source, the reconstruction fidelity slightly degraded --- we observed $F_{(nn)}= 0.86\pm0.04$ vs. $F_{(raw)}=0.81\pm0.05$, while $\pi_{(nn)}=0.84\pm0.04$ vs. $\pi_{(raw)}=0.75\pm0.07$. The most likely reason for this is some non-uniformity of the datasets caused by experimental drifts --- the data for heralded single photons were taken after some period of time. We believe, the performance may be recovered if we use heralded photons data for training as well, using a much larger amount of data. 

\subsection{State generation and detection methods}\label{app:generation}

\begin{figure}
	\center{\includegraphics[width=0.9\linewidth]{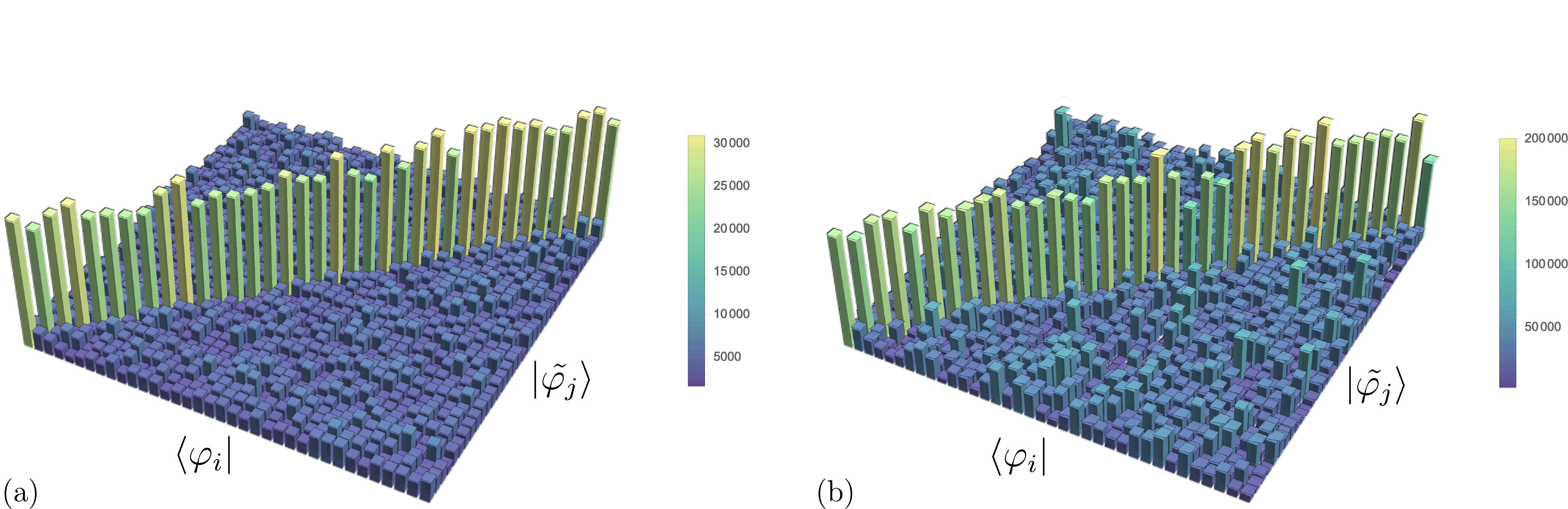}}
	\caption{Experimentally measured cross-talk probabilities $P^i_j = |\bra{\varphi_{i}} \tilde \varphi_{j} \rangle |^{2}$ for the projectors from the POVM for the cases of detection with (a) and without (b) amplitude modulation.}\label{fig:crosstalk}
\end{figure}

To generate the beams with an arbitrary phase and amplitude profiles with a phase-only SLM, we calculated hologram patterns $F(i,j)$, which can be described as a superposition of a desired phase profile $\Phi(i,j)$ and a blazed grating pattern with a period $\Lambda$ modulated by the corresponding amplitude mask $M(i,j)$:
\begin{equation}
F(i,j) = A(i,j) \mathrm{Mod}(\Phi(i,j) + 2\pi i j / \Lambda, 2 \pi),
\end{equation}
where $i$ and $j$ are the pixel coordinates. A spatially dependent blazing function allows one to control the intensity in the first diffraction order by changing the phase depth of the hologram. The presence of an amplitude mask $A(i,j)$ significantly decreases the diffraction efficiency, but at the same time corrects the alterations caused by diffraction. Bolduc et al.~showed that with the modification $\Phi(i,j) \rightarrow \Phi(i,j) - \pi A(i,j)$ this technique guarantees accurate conversion of the plane wave into a beam of arbitrary spatial profile \cite{bolduc2013exact}. This allows one to safely assume that the preparation errors in our setup are small. Considering the states generated with amplitude modulation  as ideal, we compared the quality of detection with and without such modulation ($A(i,j)=1$ for the latter case). The result is illustrated in Fig.~\ref{fig:crosstalk} where the experimentally measured probabilities $P^i_j = \Tr \left( \tilde M_j \ket{\varphi_i}\bra{\varphi_i} \right) = |\bra{\varphi_{i}} \tilde \varphi_{j} \rangle |^{2}$ are shown. The projectors $M_j = \ket{\varphi_i}\bra{\varphi_j}$ were chosen to be elements of the SIC (symmetric informationally complete) POVM for $d=6$ dimensional Hilbert space, and $\tilde M_j$ are their SPAM-corrupted counterparts. Thus the ratio between $P^i_{j=i}$ and $P^i_{j \neq i}$ was expected to be close to $36$. To quantify the deviation between $\tilde{\mathbb M}$ and $\mathbb{M}$ we used the similarity parameter $S = (\sum_{i,j} \sqrt{P^i_j \prob^i_j})^{2} / (\sum_{i,j}P^i_j \sum_{i,j} \prob^i_j)$, where $P^i_j$ and $\prob^i_j$ stand for the experimentally measured and theoretically expected probabilities, respectively. We found that the value of the similarity parameter decreased from $0.99$ to $0.96$ after switching off the amplitude modulation in the detection holograms. At the same time, the total amount of observed counts rose from $6.2 \times 10^{6}$ to $40.9 \times 10^{6}$ due to the increased diffraction efficiency of the hologram. This illustrates a known tradeoff between the projection measurement quality and detection efficiency. One of the applications of the results developed here is in increasing the detection efficiency for complex measurements of spatial states of photons without sacrificing quality.

There is a simple way to understand, why projection measurement quality is lower, than that of state preparation. The orthogonality condition for the detection of HG mode $\mathrm{HG}_{nm}(x,y)$ with a hologram corresponding to $ \mathrm{HG}_{n'm'}(x,y)$ can be written as
\begin{equation}\label{eq:overlap}
\int_{-\infty}^{\infty} \mathrm{HG}_{n'm'}^\ast(x,y)\times\mathrm{HG}_{nm}(x,y)\,dx dy = \delta_{n' n} \delta_{m' m},
\end{equation}
but since the aforementioned hologram calculation method is designed for the plain wave input and not a Gaussian beam, one has to introduce an additional Gaussian term with a waist $w_{f}$ corresponding to the detection mode waist, which breaks the orthogonality
\begin{equation}
\int_{-\infty}^{\infty} \mathrm{HG}_{n'm'}^\ast(x,y)\times\mathrm{HG}_{nm}(x,y)\times \exp[-(x^2+y^2)/w_f^2]\,dx dy \neq \delta_{n' n} \delta_{m' m}.
\end{equation}
One possible way to fix the problem is to increase the waist $w_{f}$ of the detection mode as in the experimental work \cite{bouchard2018measuring}. Unfortunately, it leads to the reduction of the detection efficiency to the level of a few percent. Thus, we used a different approach, modifying the HG mode equation \eqref{eq:HGmode} used for the holograms calculation with the second independent width parameter $\tilde{w}$ in the following way 
\begin{equation}
\tilde{\mathrm{HG}}_{a}(z) \propto H_{a}(\sqrt{2}z/w) \exp (-z^{2} / \tilde{w}^{2}),
\end{equation}
where the parameter $\tilde{w}$ was chosen to satisfy the relation
\begin{equation}
1/\tilde{w}^{2}+1/w_{f}^{2} = 1/w^{2}
\end{equation}
to compensate for the SMF term in \eqref{eq:overlap}.

\subsection{Gouy phases reconstruction by process tomography}\label{app:Gouy}

Importantly, as the lengths of optical paths in our setup were comparable with the Rayleigh lengths of the collimated beams, the generated states suffered from additional Gouy phase shifts, which depend on the mode orders. In order to avoid the related difficulties we reconstructed these phases with a standard process tomography procedure and took them into account during state generation.

Quantum process tomography is a protocol dealing with estimation of unknown quantum operation $\mathcal{E}$ acting on quantum states. The most general form of such an operation in the absence of loss is a CPTP map, which can be written in the following form
\begin{equation}
    \rho'=\mathcal{E}(\rho) = \sum\limits_{k=1}^{K} E_k \rho E_k^\dagger, \label{KrausForm}
\end{equation}
with $K\leq d^2$ in a $d$-dimensional state space, known as an operator-sum representation. The problem of quantum process tomography boils down to reconstruction of the operators $\{E_k\}$ given the observed outcomes of measurements performed on some test states $\rho_\alpha$ with probabilities 
\begin{equation}\label{eq:BornProcess}
    \mathbb{P}(\gamma|\alpha,\mathcal{E})= \Tr (M_{\alpha \gamma} \mathcal{E}(\rho_\alpha)) =
 	\Tr \Bigl(\sum\limits_{k=1}^{K}M_{\alpha \gamma} E_k \rho_\alpha E_k^\dagger \Bigr).
\end{equation}

\begin{figure}
	\center{\includegraphics[width=0.5\linewidth]{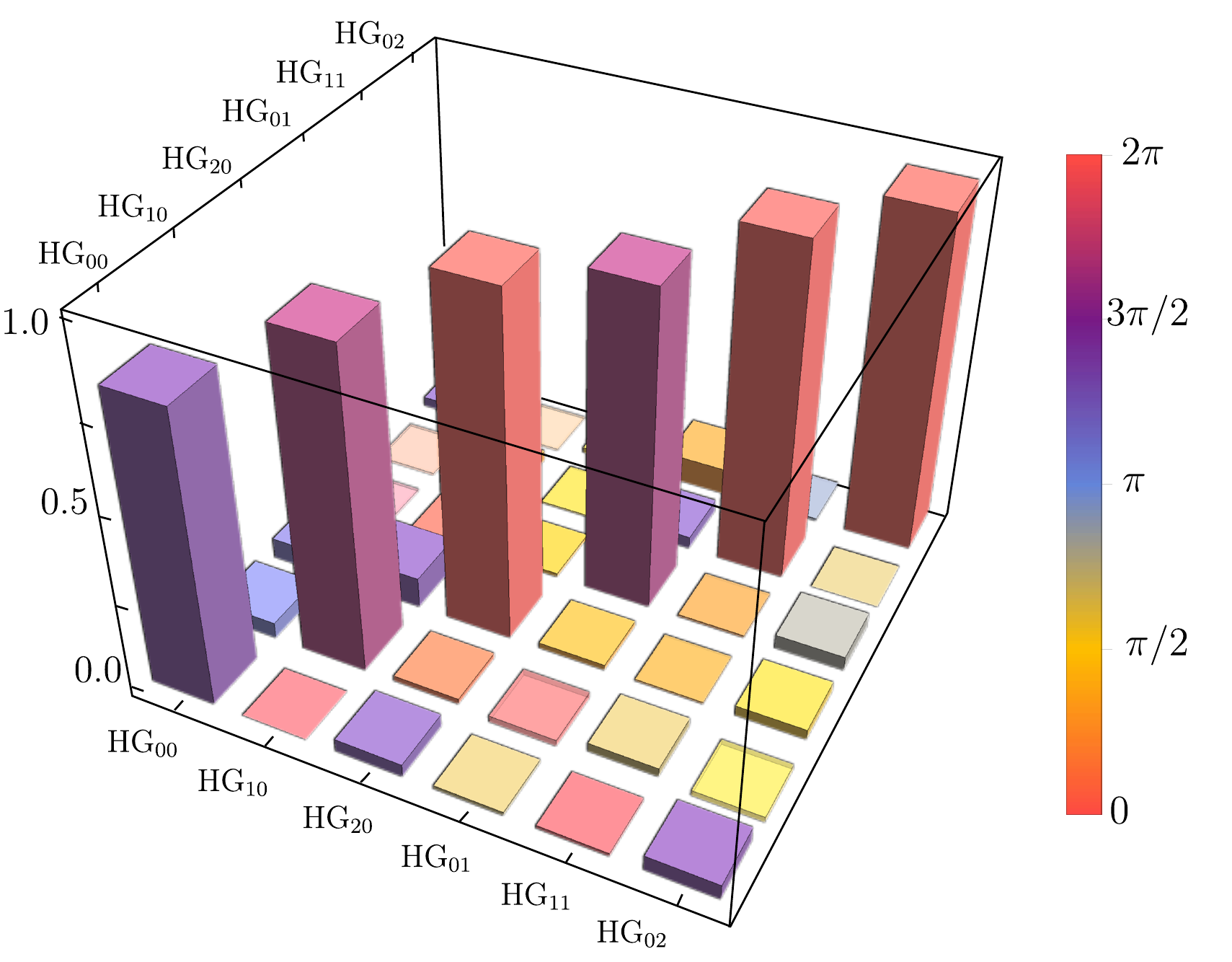}}
	\caption{Experimentally reconstructed first operator element $E_1$ of the process $\mathcal{E}$ associated with the spatial state evolution between the preparation and measurement stages. The matrix elements are expressed in Hermite-Gaussian modes basis. Ideally, it should be an identity matrix, but additional phase-shifts, known as Gouy phase-shifts were observed in our setup.}\label{fig:krauss}
\end{figure}

We have reconstructed the operator elements $E_k$ for the process associated with the spatial state evolution between the preparation and measurement. In this case masks with amplitude modulation were used both for state preparation and measurement. The process $\mathcal{E}(\rho) = \sum\limits_{k=1}^{K} E_k \rho E_k^\dagger$ turned out to be close to a rank-one process with a single dominating operator element $E_1$. As one can see from Fig.~\ref{fig:krauss}, the reconstructed $E_1$ is close to a diagonal matrix with pure phases at the diagonal. These phase-shifts are naturally interpreted as Gouy phase shifts, since they are almost equal for the modes of a particular order $n+m=\mathrm{const}$. The inferred Gouy phase shifts were found to be equal to $0.92 \pm 0.02$ and $1.97 \pm 0.03$ radians for mode orders of $n+m = 1$ and $n+m = 2$ correspondingly. Only when this additional phase-shifts were taken into account, the fidelities above 0.8 were achieved without the neural-network-enhanced post-processing.

\begin{figure}
    \subfloat[Cross-talk probabilities]
	{	
		\includegraphics[width=0.35\linewidth]{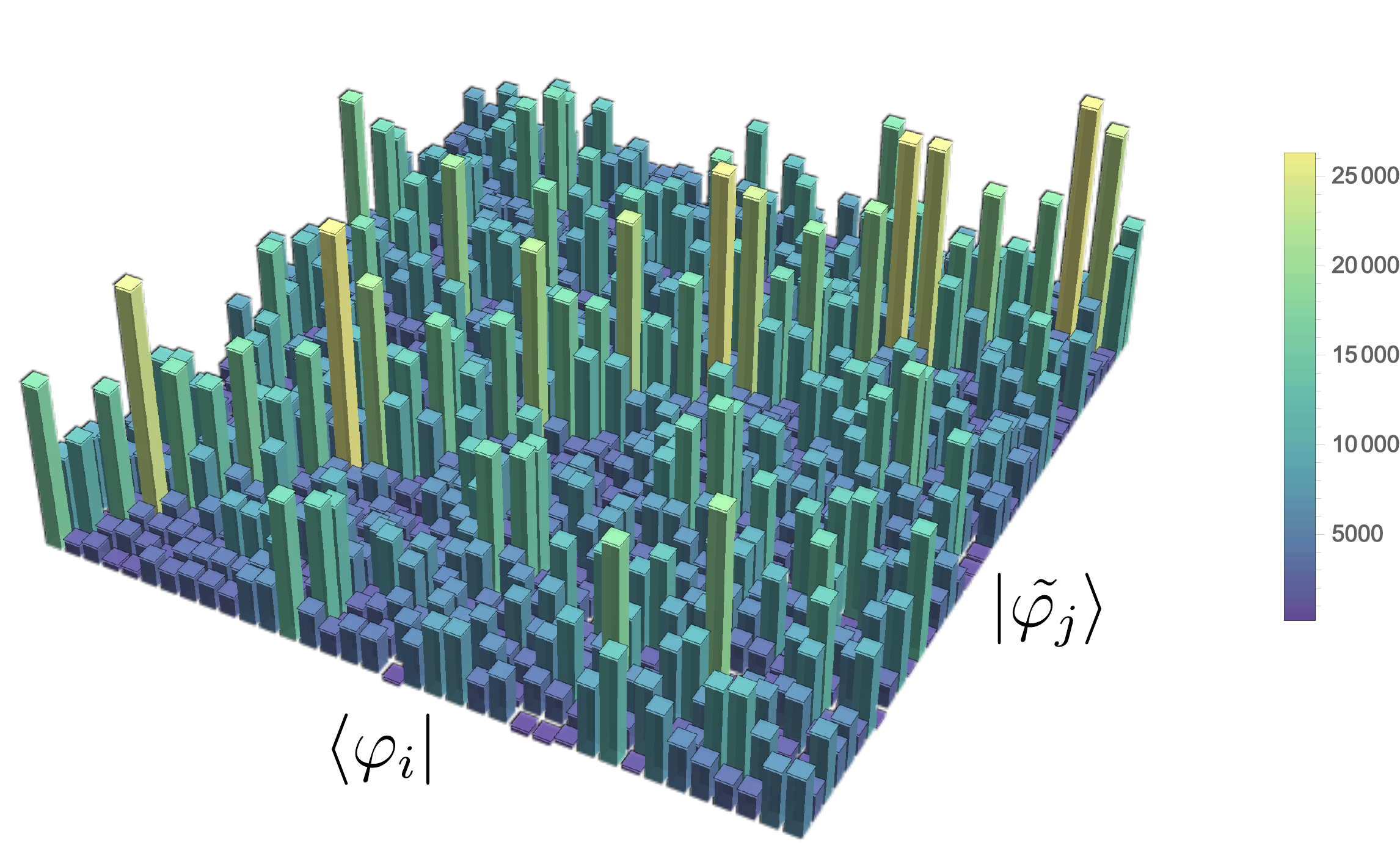}
		\label{fig:Gouy_crosstalk}
	}
    \hfill
	\subfloat[Fidelity]
	{
		\includegraphics[width=0.25\linewidth]{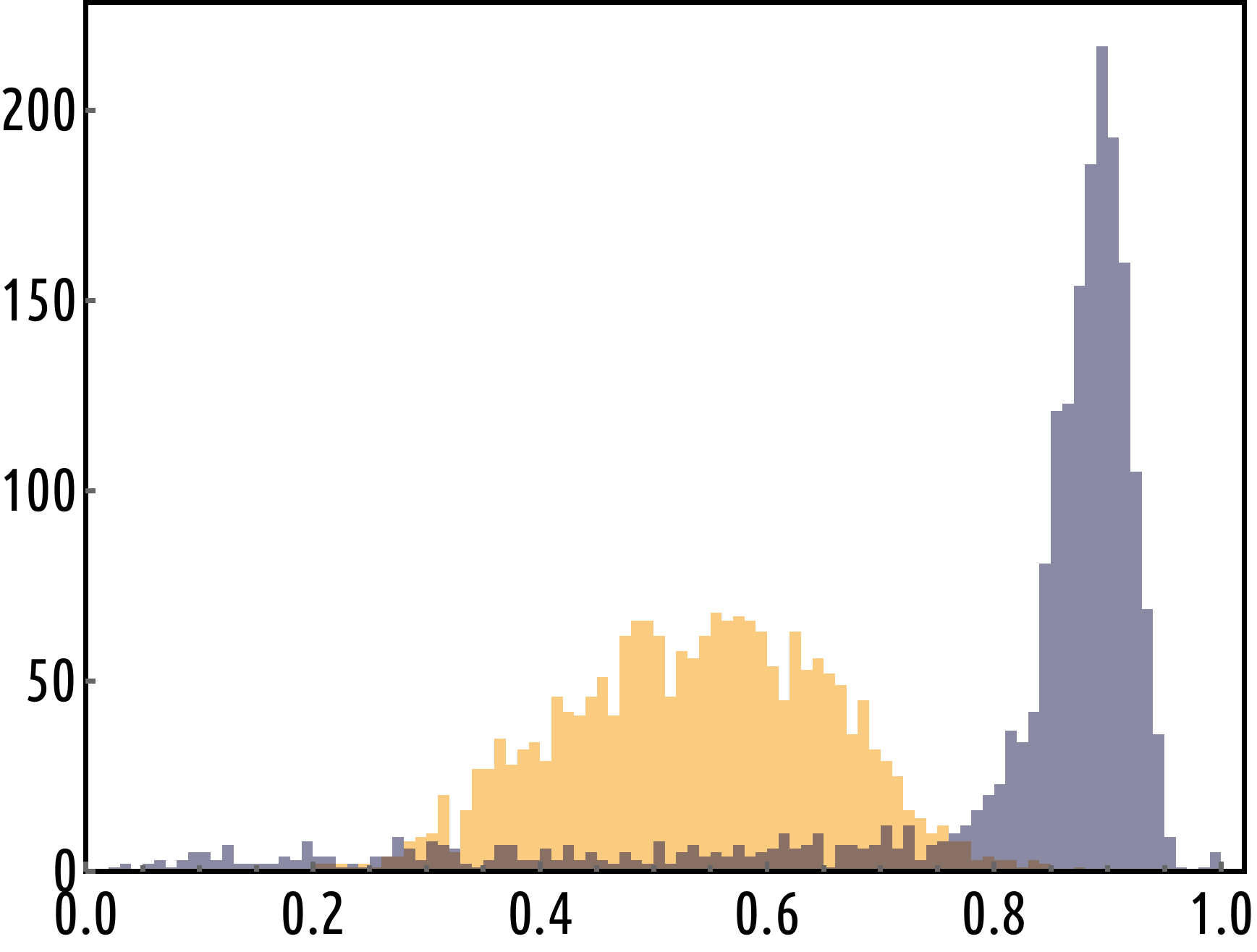}
		\label{fig:fidelityPh}
	}
	\hfill
	\subfloat[Fidelity (pure)]
	{
		\includegraphics[width=0.25\linewidth]{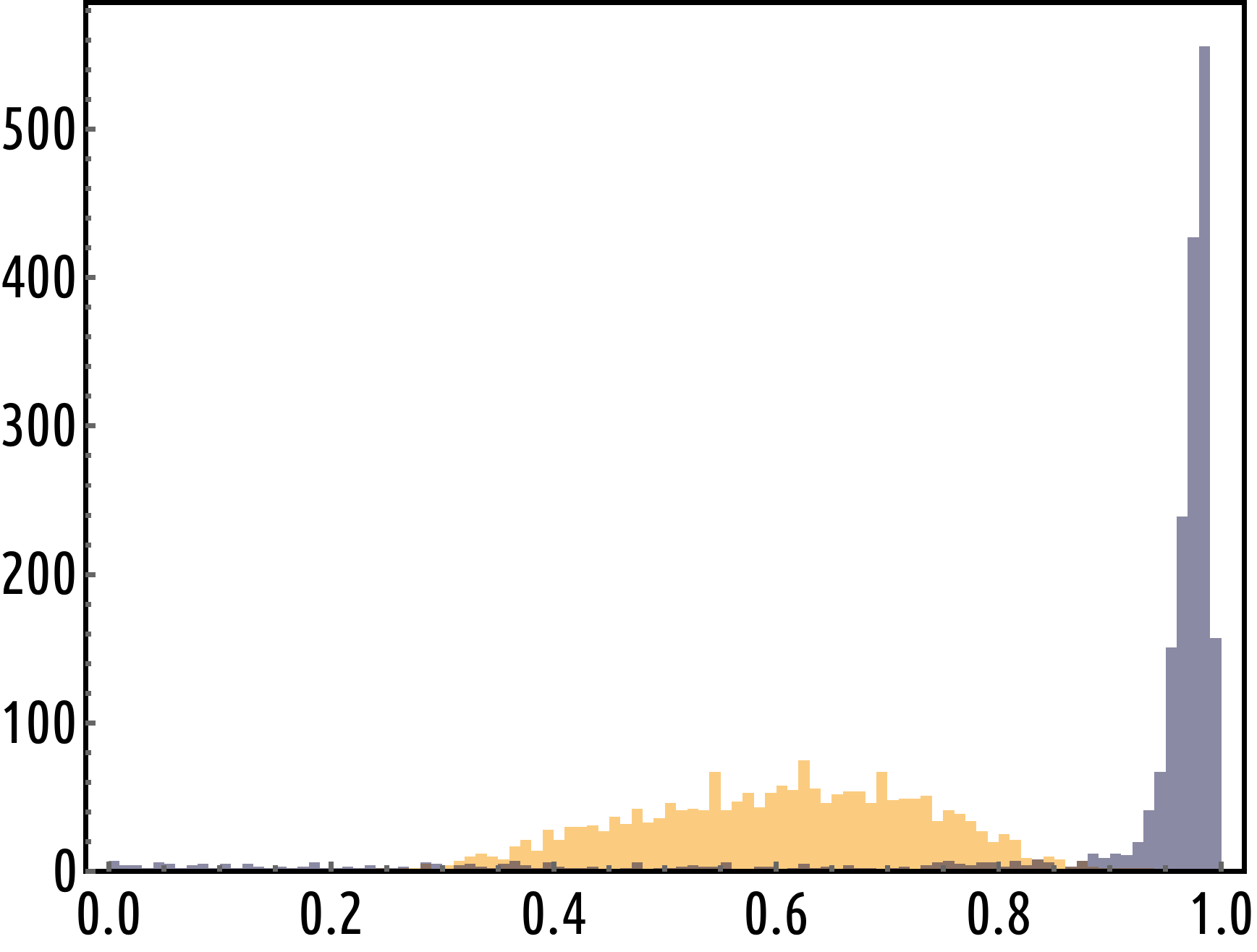}
		\label{fig:fidelityPurePh}
	}
	\caption{(a) Experimentally measured cross-talk probabilities $P^i_j = |\bra{\varphi_{i}} \tilde \varphi_{j} \rangle |^{2}$ for the projectors from the SIC POVM without the Gouy phase correction. (b) Fidelity of the experimentally reconstructed states with the ideal for 2000 test states reconstructed from the raw data (orange bars) and reconstructed after the neural network processing of the data (blue bars) without Gouy phase correction. (c) Fidelity histogram for the case of no phase correction when the states are reconstructed as pure ones.}
\end{figure}

The Gouy phase-shift elimination case is a nice example of the situation where the machine-learning-based approach helps even if the correct model of the detector is unknown to the experimenter. Indeed, instead of performing the full process reconstruction to find out the relevant phase-shifts for the modes of different order, one may stay agnostic of these shifts and consider them as just another contribution to systematic SPAM errors. We have tested the performance of the neural network trained on the states for which no correction of the Gouy phase shifts were made. The impression of how dramatically these phase-shifts affect measurement, we show the crosstalk probabilities for the projectors of the SIC POVM, with no phase correction, i.e.~modified as $\tilde{M}_j=E_1M_jE_1^\dagger$ in Fig.~\ref{fig:Gouy_crosstalk}. Without phase correction the state reconstruction of the 2000 test states gives the average fidelity of $F_{(raw)} = 0.54\pm0.12$ only and the average purity of the estimate $\pi_{(raw)}=(0.77\pm0.07)$. When the state is reconstructed as a pure one, the value of average fidelity increases to $\tilde{F}_{(raw)} = 0.60\pm0.13$. At the same time DNN-trained on the same dataset without any information about the Gouy phase-shifts gives the corresponding fidelities of $F_{(nn)}=0.81\pm0.19$ and $\tilde{F}_{(nn)}=0.89\pm0.22$ (see Fig.~\ref{fig:fidelityPh} and Fig.~\ref{fig:fidelityPurePh}).

\vskip 5pt
\subsection{State Reconstruction}\label{app:Data}

\begin{figure}[h!]
	\subfloat[Raw reconstruction]
	{	
		\includegraphics[width=0.3\linewidth]{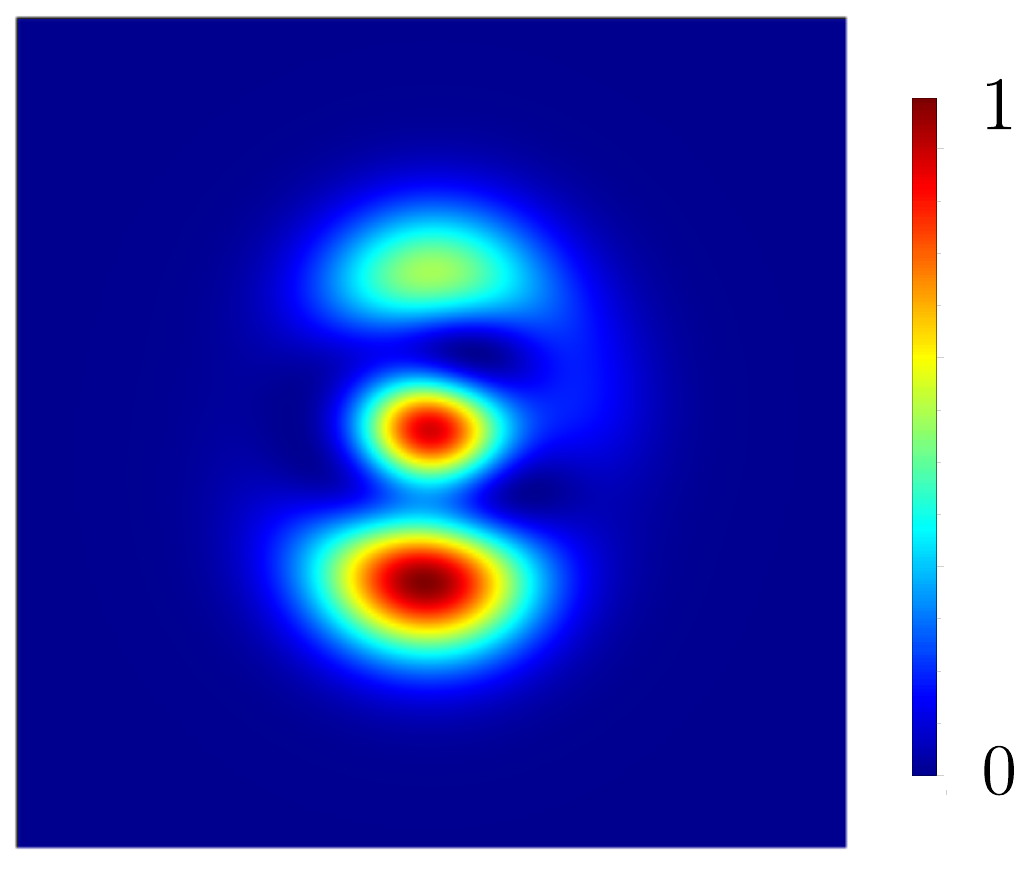}
		\label{fig:fidelity}
	}
	\subfloat[NN-enhanced reconstruction]
	{
		\includegraphics[width=0.3\linewidth]{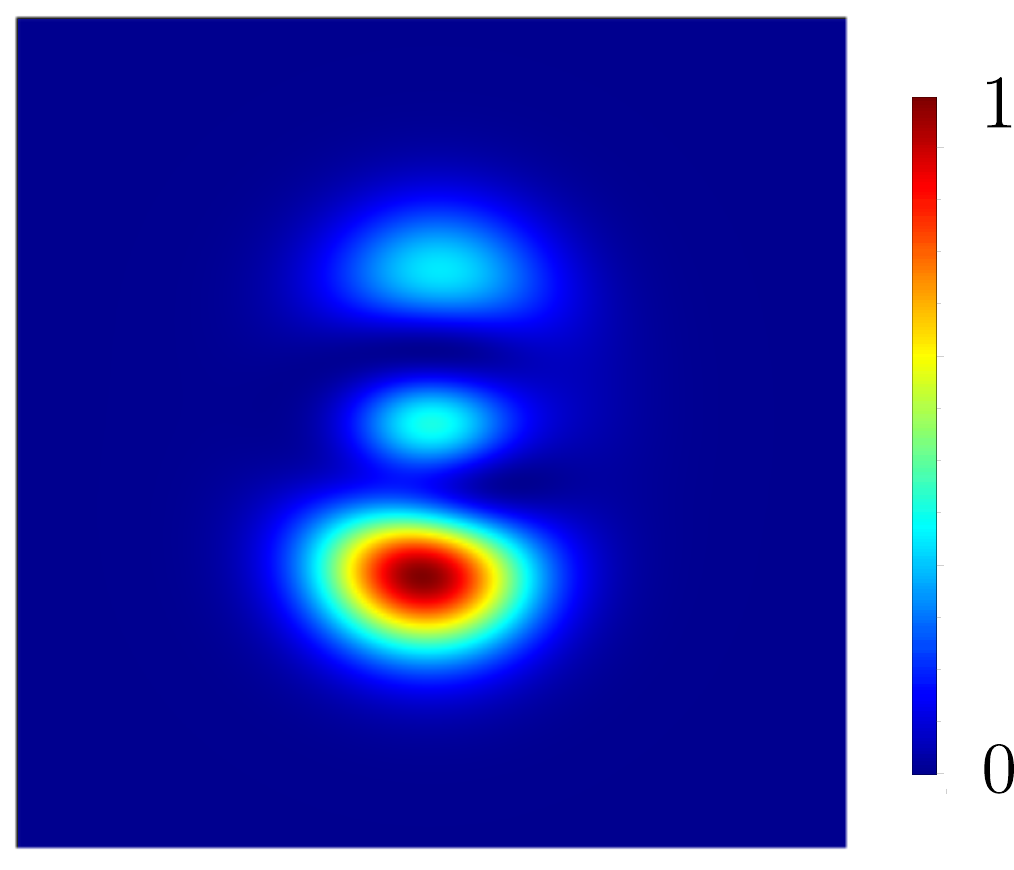}
		\label{fig:purity}
	}
	\subfloat[Prepared state]
	{
		\includegraphics[width=0.3\linewidth]{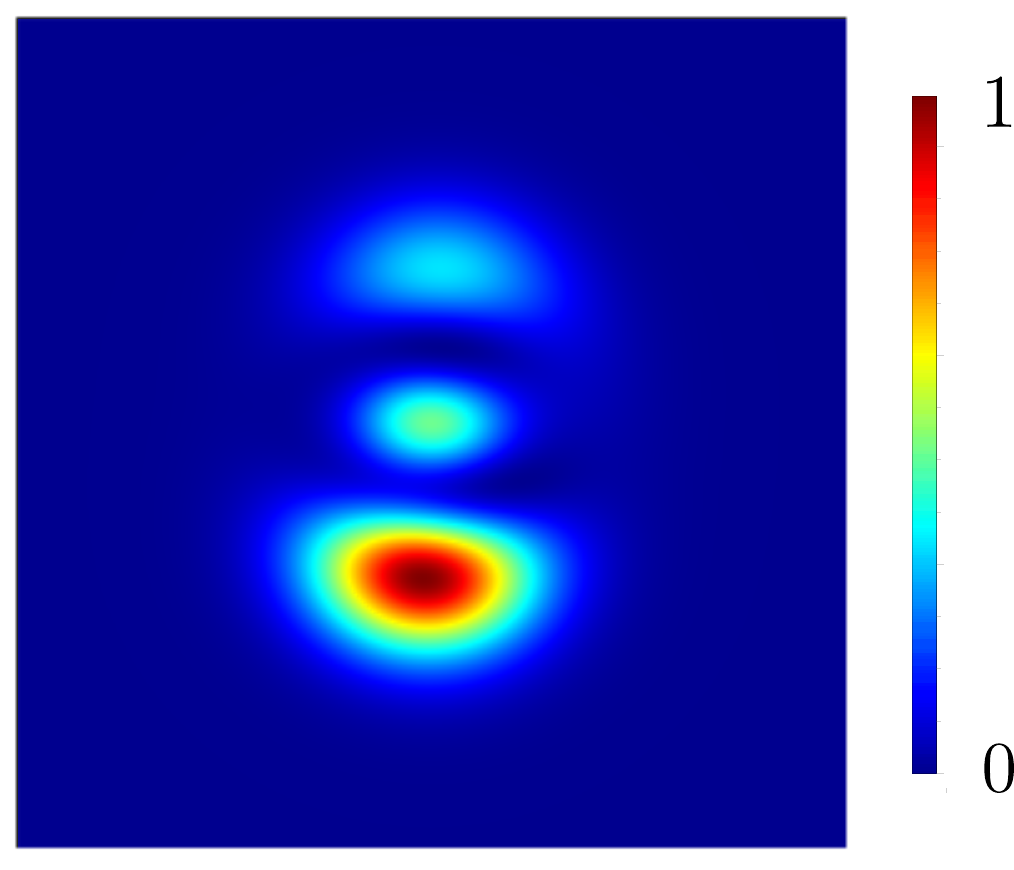}
		\label{fig:fidelityPure}
	}
	\caption{Spatial probability distributions for the exemplary reconstructed state $\ket{\psi}= (0.04 - 0.21 i)\ket{\mathrm{HG}_{00}} + (0.07 + 0.17i)\ket{\mathrm{HG}_{01}} + (-0.14 + 0.29i)\ket{\mathrm{HG}_{10}} + (0.21 - 0.05i)\ket{\mathrm{HG}_{02}}  + (0.09 + 0.02i)\ket{\mathrm{HG}_{11}} + (0.68 - 0.55i)\ket{\mathrm{HG}_{20}}$. (a) Reconstruction from the raw data, (b) reconstruction from the predicted probabilities, (c) the prepared state. Fidelities with the prepared states are $F_{(raw)}=(0.75\pm0.02)$ and $F_{(nn)}=0.91$, respectively.}
	\label{fig:ExamplaryState}
\end{figure}

\subsection{Neural Network}\label{app:nn}
Throughout the paper we consider a feed-forward neural network \cite{Goodfellow-et-al-2016} with two hidden layers of 400 and 200 neurons respectively, which maps input probabilities to the ideal ones and can be regarded as an autoencoder. To prevent overfitting in our model we use dropout between the two hidden layers with drop probability equal to 0.2; this means that at each iteration we randomly drop 20\% of the neurons of the first hidden layer in such a way that the network becomes more robust to variations in the input data. After both hidden layers we use the Rectified Linear Unit (ReLU) as activation function, while in the final output layer of 36 dimensions we use a softmax function to transform predicted values in probabilities.  
The network is trained considering the Kullback-Leibler divergence (KLD) between predicted values and the real target probabilities. Thus, we aim at minimizing the distance between the predicted distribution and the objective one according to
\begin{equation}
 	\sum_{i=1}^{N} D_{KL}(\{\prob^i\}||\{p^i\}) = \sum_{i=1}^{N} \sum_{\gamma=1}^{d^2} \prob^i_\gamma \log\left( \frac{\prob^i_\gamma}{p^i_\gamma} \right).
\end{equation}
In the following, we address the performance of DNN with respect to varying the size of the dataset, constituted by 10500 states, and the performance for the reconstruction task.

\begin{figure}
\centering
    \includegraphics[width=0.65\textwidth]{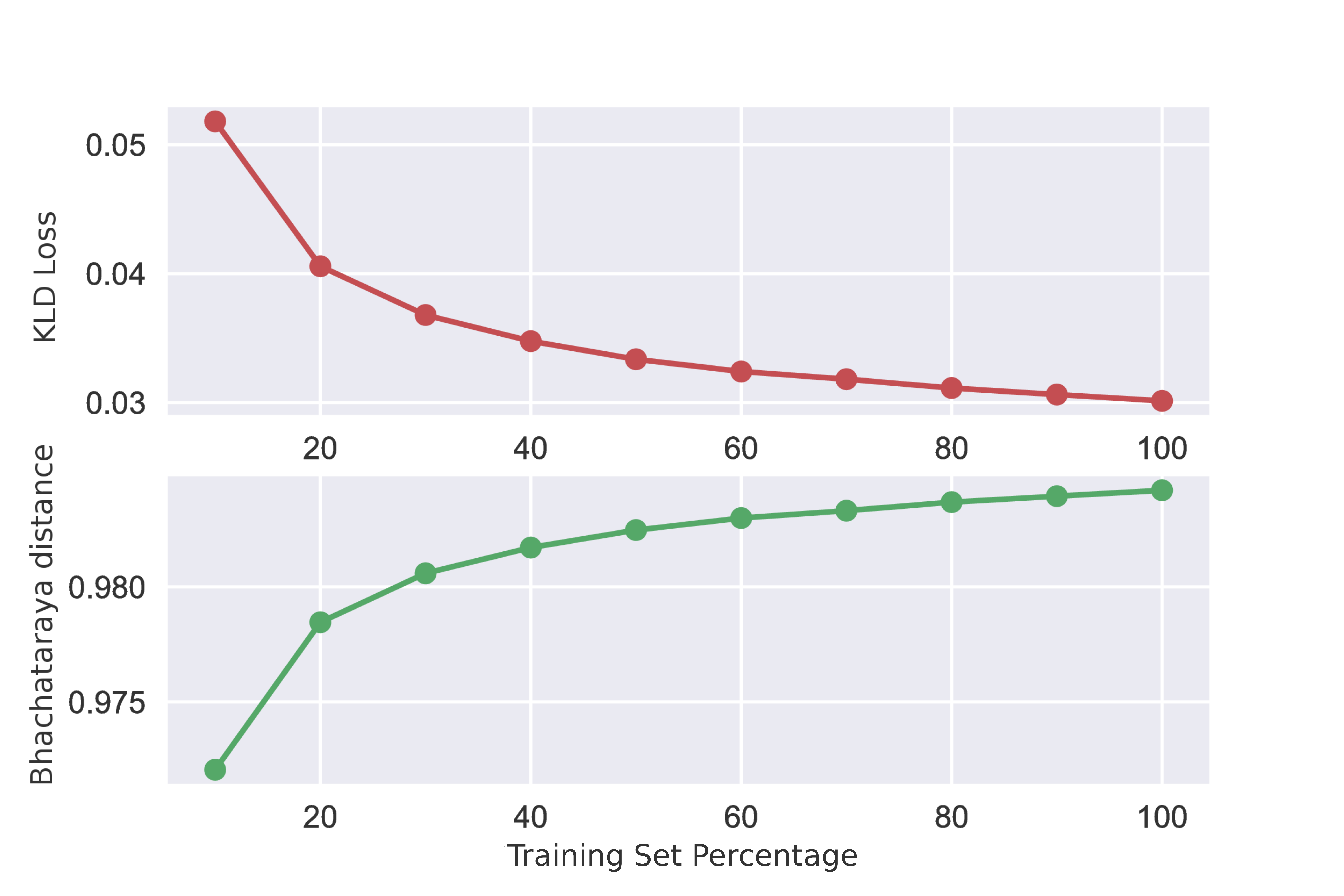}
    \caption{Dependence of the KLD loss function on the training set size (upper graph) and learning rate of the neural network quantified by classical fidelity or Bhattacharyya distance (lower graph). As expected, the  quality of the prediction offered by our model increases with increasing size of the training set.}\label{fig:training:effect}
\end{figure}

At each iteration we select 2000 states from the dataset that we consider as testing samples. Then, from the remaining dataset of $K=8500$ states we sample a percentage of data in the range from $\eta=0.1$ to $\eta=1$ with steps of 0.1 (i.e., 10\% of the data, or 850 samples). We train the network over 200 epochs and compute both the loss function and Bhattacharyya distance (classical fidelity) 
\begin{equation}
F_c(\eta)=\sum_{i=1}^{\eta K} \sqrt{\prob^i_\gamma p^i_\gamma}
\end{equation}
on the test data sample. We do this 5 times and average the results to report a stable value as shown in Fig.~\ref{fig:training:effect}. Interestingly, little data and few epochs are necessary to learn how to generate probabilities that are close to the ideal ones. Fidelity for $\eta=0.1$, i.e. training set consisting of 10\% of the data is already equal to $0.9720$. 

The experimental setting used to obtain the result described in the paper is as follows:
we divide our initial dataset into three subsets, namely training, validation and testing set, which is in line with commonly accepted ratio of 80\% (or 60\%) for training, 10\% (or 20\%) for validation and 10\% (or 20\%) for testing. We hereby approach the problem considering roughly 20\% of the dataset (2000 samples) for testing and we use 15\% as validation (1500 samples) and 7500 samples for training. This division ensures that there is enough data to train the network and that we can test on a sample that is almost 20\% of the original data. 

The training set is used to train the model while the validation set is an independent set to stop the network training as soon as the error no longer decreases on the validation set. This technique is generally referred to as \textit{early stopping} \cite{prechelt1998early}, we stop training if the error does not decrease within 100 epochs. At the end of the early stopping we restore the weights of the network that had the best validation loss during training. Finally, we test the model on the test set. This last step allows us to have a more unbiased estimation about the value of the loss on a completely unseen set of data, since the model has been chosen using a validation set and it is biased on this set.

\end{document}